\documentclass[aps, a4paper,twocolumn, 10pt, prl]{revtex4-1}
\usepackage[latin1]{inputenc}
\usepackage{lipsum}
\usepackage{amsfonts, amssymb, braket, mathtools}
\usepackage[dvipsnames, table]{xcolor}
\usepackage[caption=false]{subfig}
\usepackage[export]{adjustbox}
\usepackage{subfloat}
\usepackage{tikz}
\usepackage[capitalise]{cleveref}
\usepackage{graphicx}
\usepackage[inline, shortlabels]{enumitem}
\usepackage{bbm}
\usepackage{bbold}
\usepackage{array}
\usepackage[normalem]{ulem}
\begin{document}
\title{Correlations for computation and computation for correlations}
\author{Bülent Demirel$^{1,2}$, Weikai Weng$^{1,2}$, Christopher Thalacker$^{1,2}$, Matty Hoban$^{3}$, and Stefanie Barz$^{1,2}$}
\email{stefanie.barz@fmq.uni-stuttgart.de}
\affiliation{$^{1}$Institute for Functional Matter and Quantum Technologies, University of Stuttgart, 70569 Stuttgart, Germany}
\affiliation{$^{2}$Center for Integrated Quantum Science and Technology (IQST), University of Stuttgart, 70569 Stuttgart, Germany}
\affiliation{$^{3}$Goldsmiths, University of London, UK}

\date{\today}

\begin{abstract}
Quantum correlations are central to the foundations of quantum physics and form the basis of quantum technologies.
Here, our goal is to connect quantum correlations and computation: using quantum correlations as a resource for computation---and vice versa, using computation to test quantum correlations. 
We derive Bell-type inequalities that test the capacity of quantum states for computing Boolean functions
and experimentally investigate them using 4-photon Greenberger-Horne-Zeilinger (GHZ) states.
Further, we show how the generated states can be used to specifically compute Boolean functions -- which can be used to test and verify the non-classicality of the underlying quantum states.
The connection between quantum correlation and computability shown here has applications in quantum technologies, and is important for networked computing being performed by measurements on distributed multipartite quantum states.
 
\end{abstract}

\maketitle

\begin{figure}
	\centering
	\includegraphics[width=.75\linewidth]{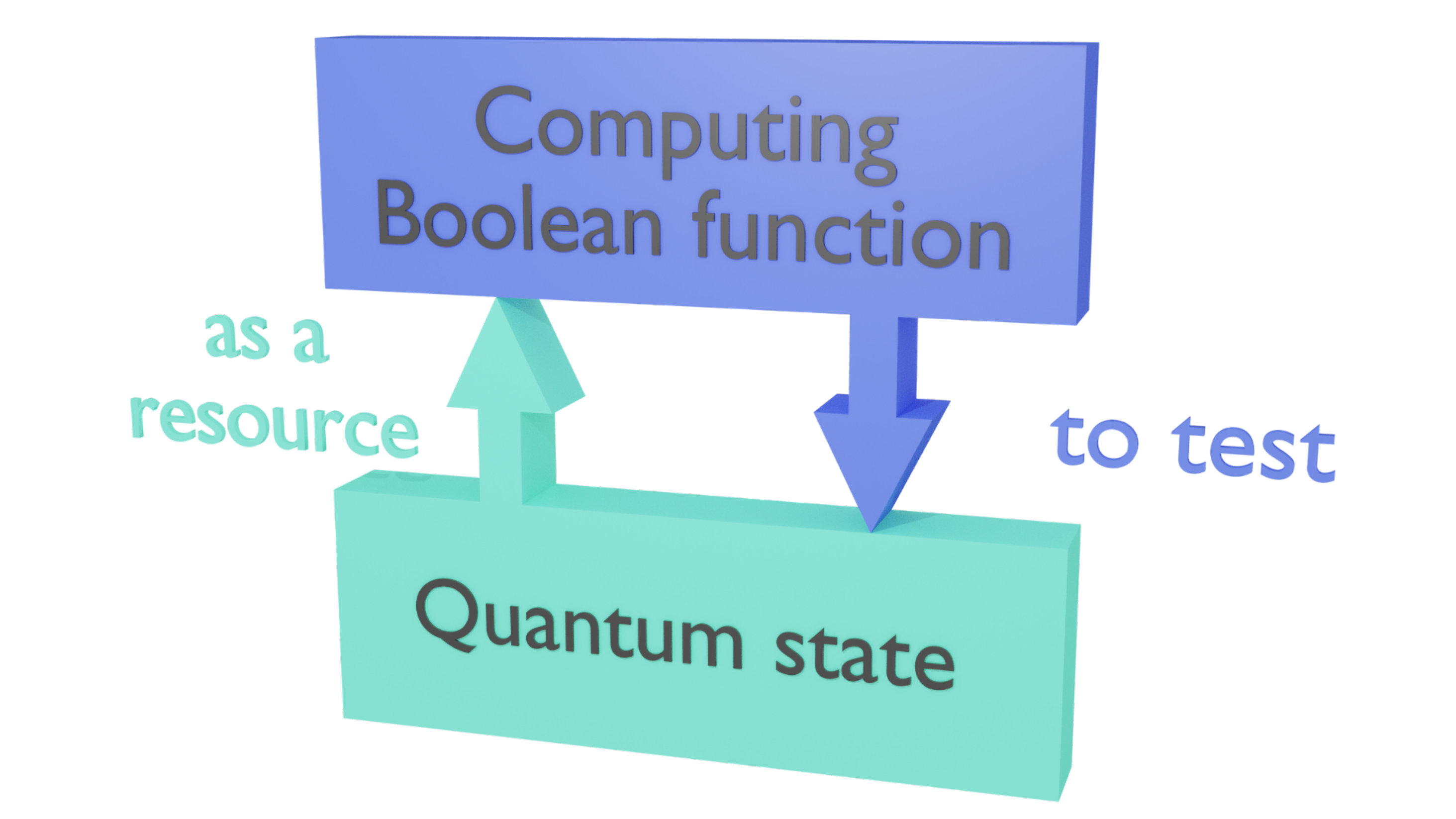}
	\caption{Illustration of the dual link between quantum states and computing Boolean functions: We can derive Bell-like inequalities to test whether certain quantum states can be used as a resource for computation. Vice versa, we can use the computation of Boolean functions as a test of quantum correlations.}
	\label{figure0}
\end{figure}

Since the beginning of quantum theory, the puzzling and non-local nature of entanglement has been a major topic of research in theoretical~\cite{Bell1964,CHSH1969,Mermin1990,Cirelson1980} and experimental physics~\cite{Aspect1982,Kwiat1995,Kurtsiefer2001,Howell2001} with the demonstration of loophole-free Bell tests being a key achievement~\cite{Hensen2015,Shalm2015, Giustina2015,Rosenfeld2017}.
Besides its fundamental nature, entanglement is one of the key ingredients of quantum technologies and forms the basis for quantum communication and quantum computing.
In quantum communication, Bell inequalities and testing correlations have practical applications and ensure the security of protocols and devices~\cite{Vazirani2019, Brunner2014}. 
Quantum computing shows speed-ups in certain computational tasks and it is believed that it will have tremendous impact~\cite{Nielsen2000}. 
Although, an advantage of quantum computers over classical computers has been shown recently for the first time~\cite{Arute2019}, current quantum devices are not yet at a stage where they can solve large-scale problems.
However, beyond full-power quantum computing, achieving an advantage in some form  of non-classical computation is highly desirable~\cite{Preskill2018}.
The main goal of this work is to demonstrate a quantum advantage in computing with simple quantum resources and to develop tools that quantify the usefulness of the resources (see Fig.~\ref{figure0}).

While the most common model of quantum computation is the circuit model, measurement-based quantum computing~\cite{Nielsen2000, Raussendorf2001}, is computationally equivalent. Here, one first generates a universal entangled quantum state and the computation is carried out by successive, \textit{adaptive} measurements on that state---measurement results are processed by a classical control and determine the settings of future measurements~\cite{Raussendorf2003}. Crucially, the classical control only needs computation with XOR and NOT gates, called \textit{linear side-processing}.
In this setting, adaptivity of the measurements is crucial: removing it disables determinism and makes universal quantum computing impossible~\cite{Shepherd2009}.

However, it has been shown that non-adaptive measurements on entangled states are a resource for universal classical computation.
For example, three-qubit GHZ states and linear computation (XOR gates) are sufficient to implement (universal) NAND gates~\cite{Anders2009}.
More generally, a linear side-processor (XOR computer) combined with non-adaptive measurements on entangled resources is sufficient to realize nonlinear Boolean functions and thus allows universal classical computation~\cite{Raussendorf2013}. 
This model is also referred to as $\text{NMQC}_{\oplus}$---non-adaptive measurement-based quantum computing with linear side-processing~\cite{Raussendorf2013}.

\begin{figure*}%
	\includegraphics[width=.99\textwidth]{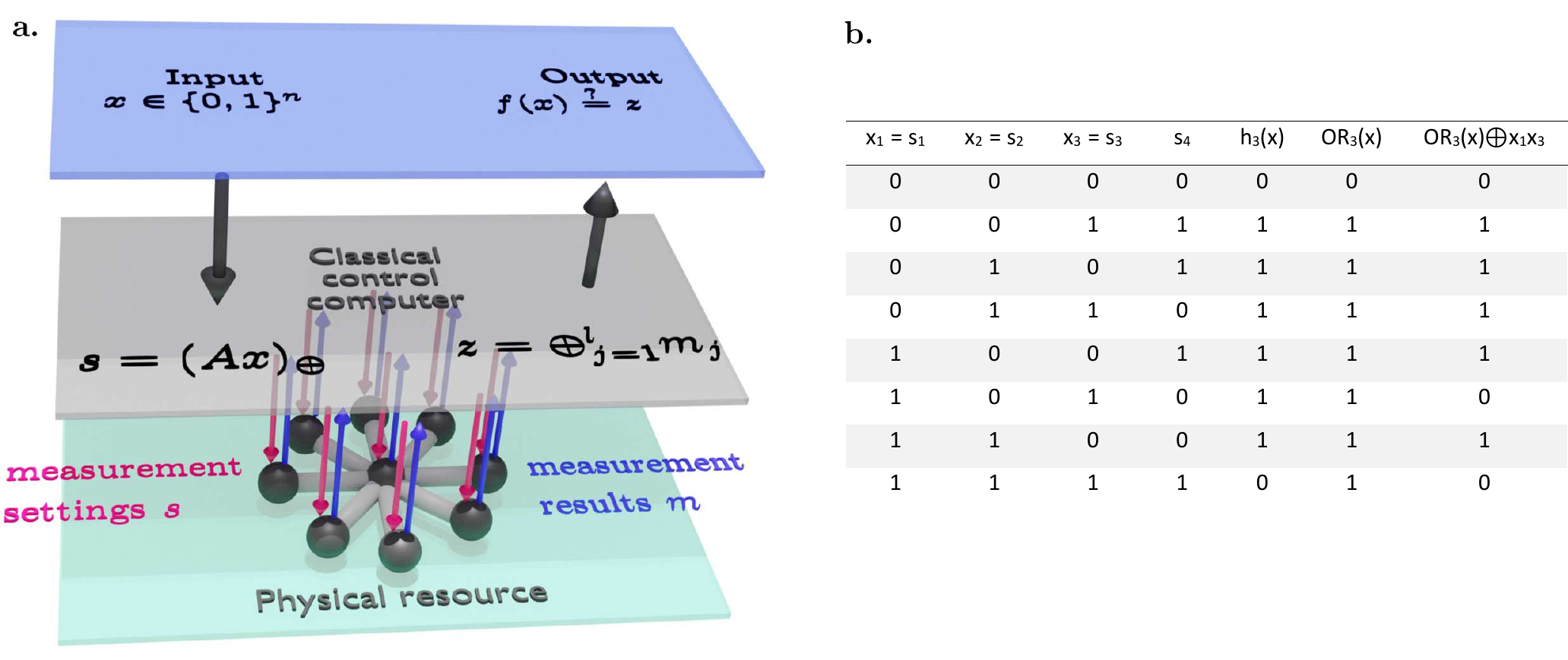}
	\caption{\textbf{a.} Concept of our setting to compute a Boolean function $f(x)$ as described in the main text.
	The input $x$ is transformed into a bit string $s$ which determines the measurement settings $M_j (s_j)$ on the physical resource.
	The outcomes of these measurements $m_j$ determine the results of the computation: $z :=\bigoplus^l_{j=1} m_j$ is generated by the parity of the outcomes from the quantum measurements. After linearly generating the bit-string $s_j$ from the input bit-string $x$, a measurement of the observables~$M_j (s_j)$ controlled by the value $s_j$ are performed, whose outcomes $m_j$ are again linearly processed to finish the computation.
	\textbf{b.} Truth table for the three functions considered in this work with input string $x =(x_1 , x_2 , x_3)$ and the bit string $s =(s_1 , s_2 , s_3, s_4)$ (see main text for details).}
	\label{figure1}
\end{figure*}


In this setting, computing a Boolean function deterministically requires a number of qubits that scales exponentially with the length of the input bit string~\cite{Hoban2011}. However, in the case of probabilistic computation of Boolean functions there is an advantage using even small-scale quantum resources.

Here, we build on this advantage and show the computation of non-linear Boolean functions with quantum resources (see Fig.~\ref{figure0}).
We link the violation of certain Bell-like inequalities to the capacity of quantum states for being a resource for computation.
We experimentally generate GHZ states, the optimal states for this task~\cite{Werner2001,Zukowski2002}, and 
 demonstrate the violation of different Bell inequalities that are related to computing certain non-trivial Boolean functions.

\section{The Setting}

The basic model of computing by non-adaptive measurements on quantum resources is shown in \cref{figure1}.
Let $x \in \{0,1\}^n$ be the input, we aim to compute the Boolean function $f: \{0,1\}^n \mapsto \{0,1\}$ (upper layer). Assume, the input $x$ is generated with probability $p(x)$.\\ 

First, the input $x$ is processed by a linear side processor with access to XOR and NOT gates (equivalent to addition modulo 2) only, as in the middle layer.
Here, the input bit-string $x$ is transformed into another bit-string $s \in \{0,1\}^l$ with $l \geq n$ and   
\begin{equation}
s_{j} = \bigoplus^n_{k=1} a_{jk} x_k
\label{eq:sj}
\end{equation}
for each $j^{\text{th}}$ bit of $s$, where $a_{jk}\in\{0,1\}$ and $\bigoplus$ is summation modulo 2. 
The values $a_{jk}$ can be seen as elements in an $l$-by-$n$ binary matrix $A$ (see \cref{figure1}\,a.) and we can write $s = (Ax)_{\oplus}$.\\

The $j^{\text{th}}$ bit $s_j$ now determines the settings for the measurement $M_j (s_j)$ on the $j^{\text{th}}$ qubit of the physical resource (bottom layer).
For each measurement~$M_j (s_j)$, we obtain a measurement outcome $m_j \in  \{0,1\}$, associated with the eigenvalues $(-1)^{m_j}$. All outcomes $m_j$ are collected in an outcome bit-string $m \in \{0,1\}^l$.
Note that the number of bits in the input $x$ is distinct from the number of parties $l$ in the physical resource. For example, in this work, we will focus on the case of $n=3$ and $l=4$.
\\

We can now pose the following question: is $z :=\bigoplus^l_{j=1} m_j$, the \textit{parity} of all outcomes, equal to $f(x)$, the designated Boolean function?
\\

To answer this question, one can determine the success probability for obtaining $z=f(x)$ to be
\begin{equation}
p(z=f(x)) = \frac{1}{2}(1+\beta).
\label{eq:SuccessProb2}
\end{equation}
with 
\begin{equation}
\beta=\sum_x p(x) (-1)^{f(x)} E(x)
\label{eqn:def_beta}
\end{equation}
being a weighted sum of expectation values $E(x) := p(z=0\vert x) - p(z=1\vert x)$.
Therefore, if $\beta=1$, then $E(x)=(-1)^{f(x)}$ for all $x$, and the function $f(x)$ can be computed \textit{deterministically}.

From Eq.~\eqref{eqn:def_beta}, we obtain a Bell-like inequality, where the upper limit that is determined by the physical resource used is
\begin{equation}
	\beta \leq  \left\{\begin{matrix} c~, & \text{classical} \\q~, & \text{quantum} \end{matrix} \right.~,
	\label{eq:GeneralBell}
\end{equation}
for classical resources (c) and quantum resources (q).  
\\

Classical resources could be simply arbitrary measurements on an n-partite separable quantum state, where the statistics are convex mixtures of local probabilities. Alternatively, we can assume a local hidden variable model~\cite{Anders2009, Hoban2011}, \textit{or} a non-contextual hidden variable model \cite{Raussendorf2013}. This can be motivated when we assume that there is no communication between the resources, and operations are local \textit{or} that local measurement on one qubit commute with local measurements on another, respectively.
The crucial point here is that all of these motivations give rise to the same experimental predictions.
Equivalently, we can assume the classical outcomes $m_j$ are solely determined by the choice $s_{j}$ and shared random variables between the parties (see Appendix for details). 
Given this notion of classical resource, it has been shown that the only functions $f(x)$ that can be computing deterministically by classical resources are \textit{linear Boolean functions}~\cite{Raussendorf2013, Hoban2011}. In this way, classical resources have the same computational power as the linear side-processor. As mentioned, quantum resources can have an advantage when the function $f(x)$ is non-linear.
\\

One can show that the maximal quantum bound $q$ is achieved by GHZ states $\ket{\text{GHZ}^{(l)}} = (\ket{0}^{\otimes l} + \ket{1}^{\otimes l})/\sqrt{2}$ and measurement of observables in the X-Y plane of the Bloch sphere $M_j (s_j) = \cos(s_j \phi_j) X +\sin(s_j \phi_j) Y $ for appropriately chosen angles~\cite{Werner2001}.
More details on the derivation of the equations above is given in the Appendix.\\

\section{Inequalities for computation.} 

\begin{figure*}\centering
	\subfloat[\label{fig:setup1}]{\includegraphics[width=0.99\textwidth]{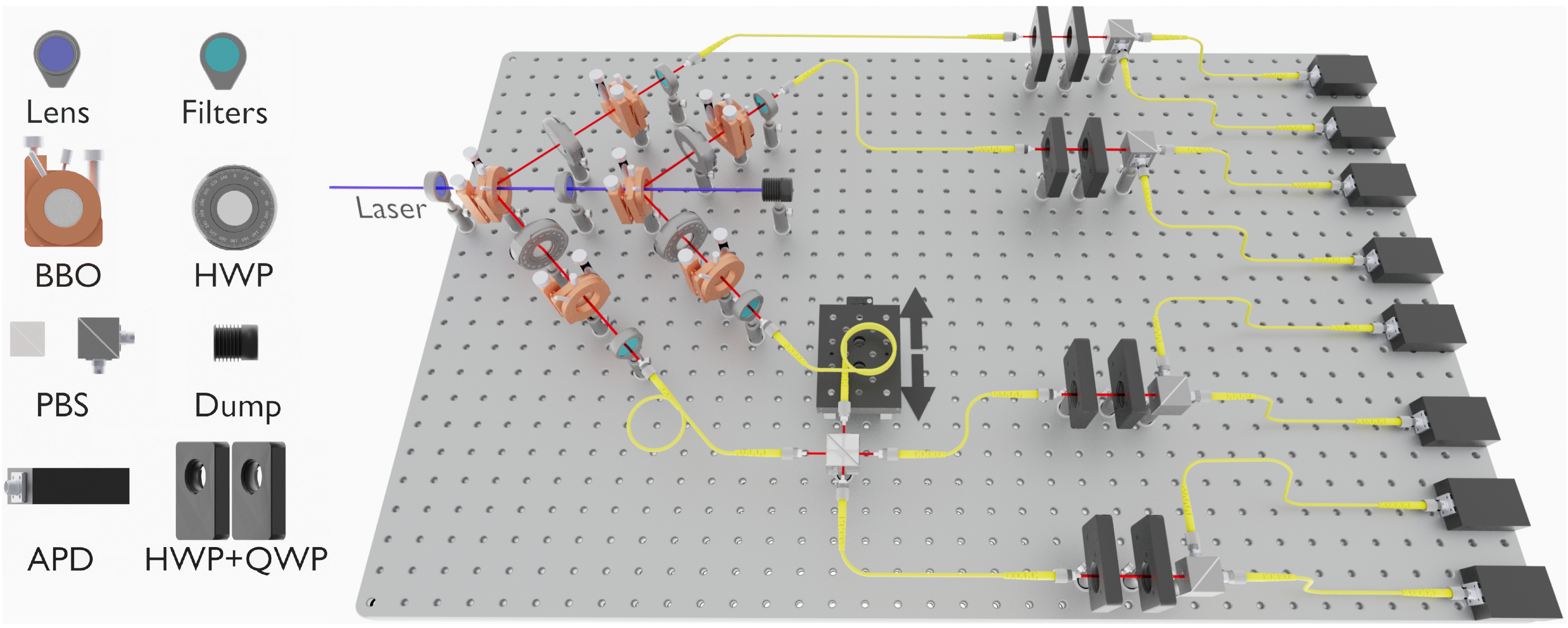}}
	\caption{ Experimental setup. A fs-pulsed Ti:sapphire laser at 780~nm is first frequency doubled and then passes through two nonlinear $\beta$-barium-borate (BBO) crystals, each of which produces spatially separated single photons by type-II spontaneous parametric down-conversion (SPDC) in states:	$\ket{\psi^{-}} = \left(\ket{H,V}-\ket{V,H}\right)/\sqrt{2}$. Half-wave plates (HWP) and additional BBO-crystals compensate walk-off effects and allow to adjust the relative phase. The photons in modes 2 and 3 of the two states $\ket{\psi_{12}^{-}} \ket{\psi_{34}^{-}}$ are sent to a polarizing beam splitter (PBS). Upon postselection to one photon in each of the output port, we obtain the state $\ket{\text{GHZ}'} = \left(\ket{H,V,V,H}-\ket{V,H,H,V}\right)/\sqrt{2}$.}
	\label{fig:setup}
\end{figure*}


We consider several functions, which are listed in \cref{figure1}\,b. 

We choose one function and derive the corresponding inequality in details; we list the results for the other functions.
\\

Let us start with the function:
\begin{equation}
h_3(x) = x_1 (x_2 \oplus x_3 \oplus 1) \oplus x_2 (x_3 \oplus 1) \oplus x_3,
\label{eqn:bf_hn}
\end{equation}
which leads us to the truth table shown in \cref{figure1}\,b and which is closely related to the n-tuple AND function in Ref.~\cite{Hoban2011}. 
First, the input bit string $x$ is transformed by a linear side processor into measurement instructions $s$
\begin{align}
 s_1 =x_1, s_2 =x_2, s_3 =x_3,
\text{ and } s_4 = x_1 \oplus x_2 \oplus x_3.
\label{eqn:sideprocess}
\end{align}
We will use this pre-processing for all examples of 3-bit Boolean functions in this work.

Now, our aim is to derive an inequality that tells us whether a certain physical resource is suitable for computing $h_3(x)$.
In order to do this, we make use of Eq.~\eqref{eqn:def_beta} and choose the uniform distribution $p(x) = 1/8$:
\begin{equation}
	 \beta_{h_3(x)}= \frac{1}{8}\left(\smashoperator[r]{\sum_{x_1=x_2=x_3}}E(x) - \smashoperator[r]{\sum_{x\setminus(x_1=x_2=x_3)}}E(x) \right) \leq  \left\{\begin{matrix} c &  \\q & \end{matrix} \right.~,
	\label{eqn:Exp1a}
\end{equation}

with $(-1)^{f(x)}$ according to the truth table in \cref{figure1}\,b. 
The maximal value of $c$ can be obtained by maximizing \cref{eqn:Exp1a} with $E(x)= E(x_1)E(x_2)E(x_3)$ and enforcing that $E(x_i)=\pm 1$ for $ i=1,2,3$.

For the quantum case measurements are made on a four-qubit GHZ state. We have that for $s_j$ equal to $0$ or $1$ the corresponding observables are given by the Pauli operators $X$ or $Y$ respectively, meaning that e.g. $(s_1, s_2, s_3, s_4) = (0,0,1,1)$ corresponds to a measurement of $XXYY$ and we obtain the inequality shown in \cref{fig:data}\,a.\\
We obtain the classical and quantum bounds:
\begin{equation}
	 \beta_{h_3(x)}\leq  1/2~\text{vs.}~1~ (c~\text{vs.}~ q).
	\label{eqn:Exp1}
\end{equation}
This means that if a physical resource violates the classical bound of this inequality, it is better suited for computing the function $h_3(x)$ than classical resources, meaning it has a higher success probability to obtain the correct result. Quantum resources can \textit{deterministically} compute this function if they have at least four qubits; for three qubits or less $l=n=3$,  the bound $q$ is equal to $1/\sqrt{2}$~\cite{Hoban2011}.
\\


Another function we consider is the three-bit OR function
 \begin{equation}
 \text{OR}_3(x) = x_1 \lor x_2 \lor x_3
\label{eqn:bf_or}
\end{equation}
 which is only~0 for $x_1=x_2= x_3= 0$ and 1 otherwise.
With the distribution $p(x=(0,0,0)) = 3/10$ and $p(x\neq(0,0,0)) = 1/10$, and measurement bases $X/Y$ as above, we obtain:
\begin{equation}
	\beta_{ \text{OR}_3(x)}\leq  4/10~\text{vs.}~ 8/10~ (c~\text{vs.}~ q)~,
	\label{eqn:Exp2}
\end{equation}
where the value for $q$ has been calculated according to \cref{fig:data}\,a.


 A similar example is the function
 \begin{equation}
\text{OR}_3(x) \oplus x_1 x_3,
\end{equation}
\label{eqn:bf_or_ext}
for which we obtain 
\begin{equation}
	\beta_{ \text{OR}_3(x) \oplus x_1 x_3}\leq  9/16~\text{vs.}~14/16~(c~\text{vs.}~ q),
	\label{eqn:Exp3}
\end{equation}
with a distribution $p(x)\in\{1/16, 3/16\}$ (see \cref{fig:data}\,a), and again, measurement observables $X$ and $Y$.
\\

Finally, we aim at computing the two-bit AND function 
 \begin{equation}
\text{NAND}_2(x)=x_1 x_2 \oplus 1. 
\label{eqn:bf_and}
\end{equation}
Choosing $s_1 =x_1$, $s_2 =x_2$, $s_3 =x_1 \oplus x_2 \oplus 1$ and $s_4 =1$ and a uniform distribution p(x), we obtain the bounds
\begin{equation}
	 \beta_{\text{NAND}_2(x)}\leq  1/2~\text{vs.}~1~(c~\text{vs.}~ q).
	\label{eqn:Exp4}
\end{equation}

This computation is equivalent to the computation of a NAND using a three-qubit GHZ state is shown in \cite{Anders2009}.
All these inequalities show that a quantum resource can violate the classical bounds for all Boolean functions considered here (see also \cref{fig:data}).
This means that the probability to compute the correct result is higher than with classical resources according to Eq.~\eqref{eq:SuccessProb2}. For details on the derivations, see Appendix.

\section{Computation for testing correlations} 
In the previous section, we used Bell-like inequalities to test whether certain physical resources are suitable for computing certain Boolean functions.
Now, we would like to use computation to probe the non-classicality of the resource state.
In other words, we perform computation in our model (see \cref{figure0}) and if we obtain the correct result with a certain probability, given by the inequalities above, we know our resource has to be non-classical in a particular, formal way.

Using Eq.~\eqref{eq:SuccessProb2} we can convert the classical and quantum bounds above into success probabilities
\begin{align}
&h_3(x):&	0.750~&\text{vs.}~1.000 \label{eqn:prob_1}\\
&\text{OR}_3(x):&	0.700~&\text{vs.}~0.900 \label{eqn:prob_2}\\
&\text{OR}_3(x) \oplus x_1 x_3:&	0.813~&\text{vs.}~0.938 \label{eqn:prob_3} \\
&{\text{NAND}}_2(x):& 0.500~&\text{vs.}~1.000.  \label{eqn:prob_4}
\end{align}
Here, the first value in each row indicates the maximum probability to obtain the correct results when the function is computed using classical resources, the second value indicates the probability for computing with quantum resources.

If we perform computation in our model and achieve a success probability higher than the classical probabilities given in \crefrange{eqn:prob_1}{eqn:prob_4},  we know that our resource state is non-classical. In particular, we see that the deterministic computation of non-linear Boolean functions is a signature of non-classicality.

\section{Experiment} 

\begin{figure*}
	
	\includegraphics[width=.99\textwidth]{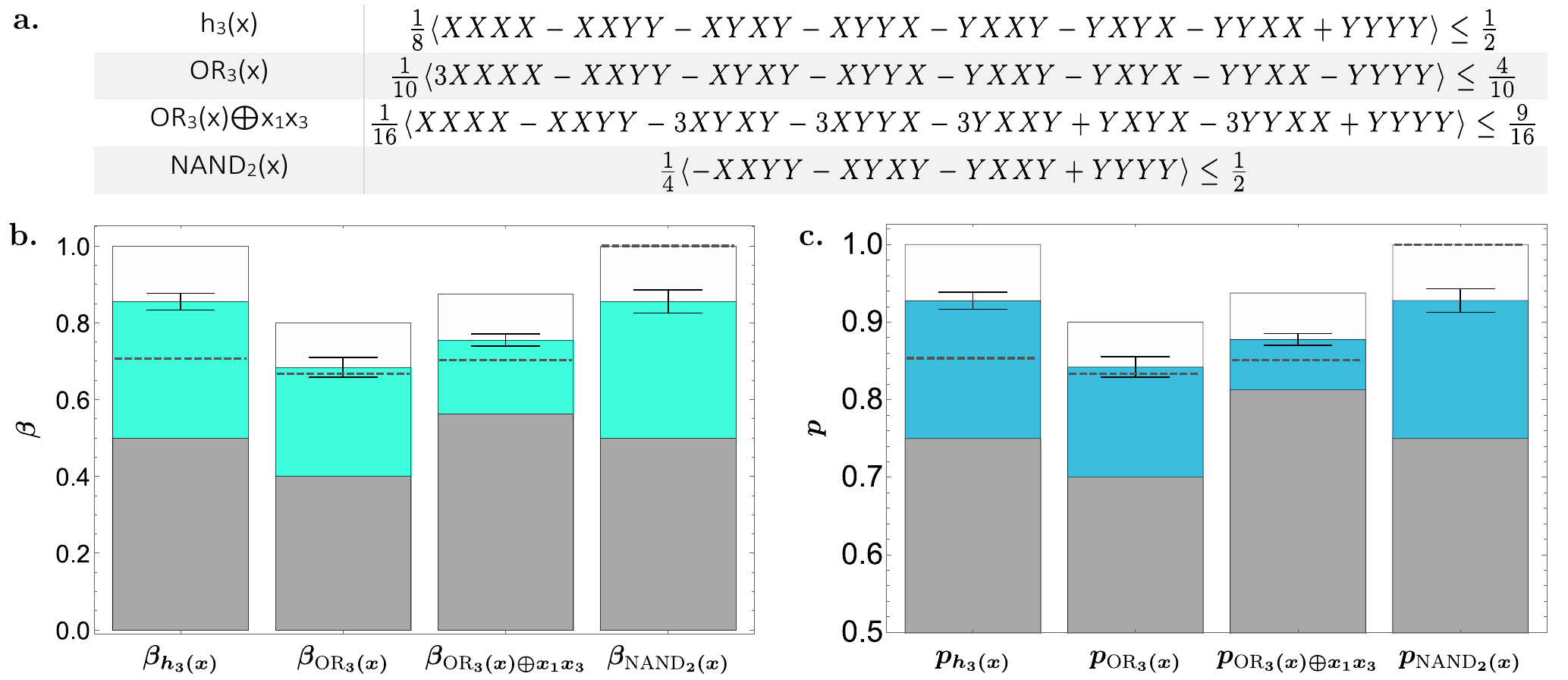}
		\caption{\textbf{a.} Overview of the tested multi-partite Bell inequalities with their corresponding Boolean functions. \textbf{b - c.} Classical, quantum, and experimentally obtained bounds for computing Boolean functions shown in the panels below.	For both the bounds in \cref{fig:data}\,b. and the probabilities \cref{fig:data}\,c. the gray area of the bars indicate the regions completely obtainable with linear operations on classical resources. The colored regions with the measurement points on top indicate that these limits have been surpassed in each case by $\beta_{h_3(x)}: 16\sigma$, $\beta_{ \text{OR}_3(x)}: 10\sigma$, $\beta_{ \text{OR}_3(x) \oplus x_1 x_3}: 8\sigma$ and , $\beta_{\text{NAND}_2(x)}: 11\sigma$.  Ergo, the correct computation of the non-linear Boolean function is more probable with quantum resources. The underlying white bars indicate the optimal quantum limits for 4 qubit measurements, while the gray dashed lines mark the quantum bound for three-qubits, hence we see that with increased entangled resources the limit moves away form the classical bound unless it is already deterministic.}
	\label{fig:data}
\end{figure*}

For exploring relation between computation and Bell inequalities experimentally, we generate four-photon GHZ states using an all-optical setup that is shown and described in \cref{fig:setup}.
The state we obtain in our experiment is 
 \begin{equation}
\ket{\text{GHZ}'} = \left(\ket{H,V,V,H} - \ket{V,H,H,V}\right)/\sqrt{2}
\label{eqn:ghzexperiment}
\end{equation} 
with $\ket{H} \mathrel{\widehat{=}} \ket{0}$ and $\ket{V} \mathrel{\widehat{=}} \ket{1}$ denoting horizontal and vertical polarisation.
Note that the state $\ket{\text{GHZ'}}$ is related to state $\ket{\text{GHZ}^{(4)}}$ by local unitary transformations, e.g. $\ket{\text{GHZ}^{(4)}} = \mathbb{1}XXZ \ket{\text{GHZ'}}$. 
We verify the state obtained in the experiment through quantum state tomography~\cite{James2001}.
The reconstructed density matrix~$\rho_{exp}$ shows a fidelity $F = \braket{\text{GHZ}'\vert \rho_{exp} \vert \text{GHZ}' }$ of ~$F = 0.82 \pm 0.01$ (see Appendix). 

The values of $\beta^{Exp}$ we obtain for the individual Boolean functions are listed in \cref{fig:data}\,b, together with the classical and quantum bounds.
All values are clearly above the classical limit by more than 8 standard deviations in the least.

If we in turn use the GHZ state generated in the experiment to perform computation, we can quantify the probability to get the correct output.
The corresponding probabilities are shown in \cref{fig:data}\,c. 
This confirms that, for all probabilities, we lie above the classical values, which verifies that our physical resource must be quantum. In addition the gray dashed line in \cref{fig:data} highlights the limits of the computation on 3 qubits.

The discrepancy to the quantum bounds arises due to experimental imperfections.
First of all, our resource state is not perfect.  The quality of the 4-photon entanglement is limited by
purity of the generated two-photon entangled states (two photon fidelity $F \cong 0.96$) as well as by the interference of the photons in modes $2$ and $3$ where we measured a Hong-Ou-Mandel dip visibility of $V =0.80 \pm 0.02$.
In addition, imperfections in the polarisation states, polarisation drifts, and higher-order emissions ($1.2 \pm 0.1 \%$ of the coincidence rate for each SPDC) reduce the quality of the generated GHZ state.




\section{Conclusion}
In this work, we link a deeply fundamental question---the violation of a Bell inequality---to computing classical functions.
We investigate this connection from two angles: verifying correlations through Bell tests quantifies the ability of a certain physical resource for computation.
Furthermore, doing computation can be used as a tool to test non-classicality itself. 
We demonstrate this connection in a quantum optics experiment and show that already a four-qubit quantum states can provide an advantage.

The beauty of this connection between classicality and linear Boolean functions is that no matter how large the classical resource, its computational power does not change. However, as we increase the number of qubits in a quantum resource the computational power increases.

An interesting question is to study further types of classical resources in our setting.
We could, for example, allow communication between measurement sites, for example, as in studies of non-locality~\cite{Gallego2012, Bancal2013}.
Given these or other additional powers, the question is how the success probabilities of the enhanced classical resources compare to quantum resources.
In addition, what do the computed Boolean functions imply about the amount of non-classicality of the resource? 

The relation between non-classicality and computing investigated here is related to the connection of Bell inequalities and quantum games~\cite{Broadbent2008}.
It is also related to work on contextuality and the use of single-qubit operations for classical computation~\cite{Dunjko2016, Barz2016, Clementi2017, Mansfield2018}.

Even, if no fully fledged quantum computer is available, our work demonstrates the advantages of quantum resources for computation.
In particular, our work has implications for quantum networks. Although, our approach here has been computational and not cryptographic, the quantum advantage in our work can be applied to a cryptographic setting if the shared resource state is distributed among agents in a network. For example, our methods could be directly used to self-test GHZ states~\cite{MillerShi2013} and generate randomness~\cite{delaTorre2015}, both in a device-independent manner.
Furthermore our quantum advantages for computation can be turned into an advantage for communication complexity~\cite{Brukner2004}.
Thus, our work is a further example how the power of modern quantum technologies lies in fundamental quantum physics.

\section{Acknowledgments}

We thank Akshey Kumar and Katharina Stütz for setting up the early stages of the experiment, and Benedikt Burichter for assisting with software.

We acknowledge support from the Carl Zeiss Foundation, the Centre for Integrated Quantum Science and Technology (IQ$^\text{ST}$), the German Research Foundation (DFG), the Federal Ministry of Education and Research (BMBF, project SiSiQ), and the Federal Ministry for Economic Affairs and Energy (BMWi, project PlanQK).


%

\onecolumngrid
\appendix
\section{Appendix}

In the following sections we give some more details on the derivation of the different multi-partite Bell-inequalities presented in the main manuscript and show a more detailed data analysis. For a more general discussion of the theory we refer to~\cite{Hoban2011}. First we outline what we mean by classical resources.

\section{Definition of classical resources}
Here, we formalise the notion of non-classicality as used in the main manuscript.
\\
A classical resource means that the probabilities (correlations) $p(m|s)$ of getting outcomes $m:(m_{1},...,m_{l})$ given measurement choices $s$ can be written as 
\begin{equation}
p(m|s)=\sum_{\lambda}p(\lambda)\prod_{j=1}^{l}\delta_{m_{j},\mu(s_{j},\lambda)}, 
\end{equation}
where $\{\lambda\}$ is a set of shared random variables with probability distribution $\{p(\lambda)\}_{\lambda}$ and $\mu(s_{j},\lambda)\in\{0,1\}$ is map from the measurement choice $s_{j}$ and $\lambda$ to a bit-value. Since the string $s$ and bit-value $z$ result from linear computation on $x$ and $m$ respectively, it can be seen that $p(z|x)$ will only be a mixture delta functions $\delta_{z,h(x)}$, where $h(x)$ is a linear Boolean function in $x$.
When we come to find the optimal classical bounds of the inequalities below, by convexity we only need to consider deterministic correlations, e.g. $p(m|s)=\prod_{j=1}^{l}\delta_{m_{j},r_{j}s_{j}}$, where $r_{j}\in\{0,1\}$ for all $j$. More formally, the set of classical correlations is a convex set and the optimal value of an inequality will be given by extreme points of the set. These extreme points are the deterministic correlations.\\

Such a classical model as above can motivated in many ways: since, in the quantum case, there is no communication between the resources, and operations are local, a local hidden variable model is the natural classical analogue \cite{Hoban2011}; since a local measurement on one qubit commutes with a local measurement on another, this motivates a non-contextual hidden variable model \cite{Raussendorf2013}. 
Furthermore, if we associate a classical resource state with a separable quantum state, then the statistics will be convex mixtures of local probabilities. 

\section{Derivation of the inequalities}

In the following, we show in detail how to derive the Bell inequalities for the Boolean functions $f: x^n \mapsto x$ given in the main manuscript:
\begin{equation}
\begin{split}
\beta = \sum_{x}\beta(x)E(x) = \sum_{x}p(x)(-1)^{f(x)}E(x) \leq \left\{\begin{matrix} c~, & \text{classic} \\q~, & \text{quantum} \end{matrix}  \right.~.
\end{split}
\label{eq:SuccessProb2SI}
\end{equation}

\subsection{The function $f(x) = x_1 (x_2 \oplus x_3 \oplus 1) \oplus x_2 (x_3 \oplus 1) \oplus x_3$}
The first example is the function $f(x) = x_1 (x_2 \oplus x_3 \oplus 1) \oplus x_2 (x_3 \oplus 1) \oplus x_3$ and all input strings $x$ are uniformly distributed $p(x) = 2^{-3}$. The function satisfies $f(0,0,0) = f(1,1,1) = 0$, else it will yield the result 1 as shown in the Table of Fig.~2 of the main section. Accordingly, from \cref{eq:SuccessProb2SI} we get the relation
\begin{equation}
	\beta = \frac{1}{8}\left( E(0,0,0) - E(0,0,1)- E(0,1,0) - E(0,1,1)- E(1,0,0) - E(1,0,1) - E(1,1,0)+ E(1,1,1)\right)~.
	\label{eq:beta1}
\end{equation}

For computing the largest classical bound, we maximize the sum in \cref{eq:beta1} over all possible values with $E(x_1, x_2, x_3)= E(x_1)E(x_2)E(x_3)$ and the expectation values confined to $E(x_i) = (\pm 1)$ for $i \in \{1,2,3\}$.

In \cref{eq:beta1} we obtain a maximal classical value of $\frac{1}{2}$.

The value for the quantum bound depends on the number of sites $l\geq n$. In our measurements we have $l=4$ and the measurement choices are encoded by the following linear map
\begin{equation}
\left(\begin{array}{c} s_1\\s_2\\s_3\\s_4\end{array} \right) = \begin{pmatrix}
1 & 0 & 0 \\ 0 & 1 & 0 \\ 0 & 0 & 1 \\ 1 & 1 & 1   
\end{pmatrix}\left(\begin{array}{c} x_1\\x_2\\x_3\end{array} \right)_{\oplus}~.
\end{equation}
Therefore, choosing for $s = 0$ observable $X$ and for $s = 1$ observable $Y$ the inequality \cref{eq:beta1} becomes
\begin{equation}
\frac{1}{8} \langle XXXX - XXYY - XYXY - XYYX  -  YXXY - YXYX - YYXX + YYYY\rangle \leq 1 \label{eq:Ineq1}~.
\end{equation}
The maximum in \cref{eq:Ineq1} is obtained exactly by $\ket{\text{GHZ}^{(4)}} = (\ket{0,0,0,0}+\ket{1,1,1,1})/\sqrt{2}$, naturally the values of all other states will be within the region bounded by this GHZ state. Note that this value of $1$ for the quantum bound is the maximum allowed algebraically.

\subsection{The function $f(x) = \text{OR}_3(x)$}

The second example is the OR function. At this point it is also worthwhile to rewrite the function in algebraic normal form (ANF) so it can obviously be seen as non-linear:
\begin{equation}
\text{OR}_3(x)=x_{1}x_{2}x_{3}\oplus x_{1}x_{2} \oplus x_{1}x_{3} \oplus x_{2}x_{3} \oplus x_1 \oplus x_2 \oplus x_3.
\end{equation}
The distribution for this function is chosen as $p(x) = \frac{1}{10}$, except for $p(0,0,0) = \frac{3}{10}$. We thus obtain the inequality
\begin{equation}
	\beta = \frac{3}{10} E(0,0,0) - \frac{1}{10} \sum_{x\neq (0,0,0)}E(x).
\label{eq:beta2}
\end{equation}
In the same way as above we can compute the classical bound to be $\frac{4}{10}$
and the quantum bound
\begin{equation}
 \frac{3}{10}\langle XXXX \rangle - \frac{1}{10}\langle XXYY + XYXY + XYYX + YXXY + YXYX + YYXX + YYYY\rangle \leq \label{eq:Ineq2}\frac{8}{10}~.
\end{equation}
This bound can be readily confirmed to be the maximum allowed for all possible quantum resources (both states and measurements) using the methods described by Werner, Wolf, \.{Z}ukowski and Brukner \cite{Werner2001,Zukowski2002}.

\subsection{The function $f(x) = \text{OR}_3(x) \oplus x_1 x_3$}
The third example is $f(x) = \text{OR}_3(x) \oplus x_1 x_3$ and the distribution $p(0,0,0) = p(0,0,1) = p(1,0,1) = p(1,1,1) = \frac{1}{16}$ and $p(0,1,0) = p(0,1,1) = p(1,0,0) = p(1,1,0) = \frac{3}{16}$. It should be noted that this function is still clearly non-linear after converting it into ANF using the identity described above. The correct signs can be read off from the truth table and we get
\begin{equation}
\beta = \frac{1}{16}\left( E(0,0,0)- E(0,0,1) + E(1,0,1) + E(1,1,1)\right) -\frac{3}{16}\left(E(0,1,0) - E(0,1,1) - E(1,0,0)+ E(1,1,0) \right) ~.
\label{eq:beta3}
\end{equation}
with a classical bound of $\frac{9}{16}$
\begin{equation}
\frac{1}{16}\langle XXXX - XXYY+ YXYX + YYYY\rangle - \frac{3}{16}\langle XYXY + XYYX + YXXY + YYXX \rangle \leq \frac{14}{16}\label{eq:Ineq3}~.
\end{equation}
Again, this bound can be readily confirmed to be the maximum for all quantum resources  (both states and measurements) using the methods described by Werner, Wolf, \.{Z}ukowski and Brukner \cite{Werner2001,Zukowski2002}.

\subsection{The function $\text{NAND}_2(x)$}

In the work of Anders and Browne \cite{Anders2009}, it was demonstrated that a three-qubit GHZ state can be used to compute the NAND function of two bits, which we can write as $\text{NAND}_2(x)=x_{1}x_{2}\oplus 1$. Since this is just a function on two bits, things will be somewhat simplified. The distribution over these two bits is $2^{-2}$ for all values of $x:=(x_{1},x_{2})$. The inequality can be written as:
\begin{equation}
\beta = \frac{1}{4}\left(-E(0,0)- E(0,1) - E(1,0) + E(1,1)\right) ~.
\label{eq:beta4}
\end{equation}
The classical bound for this inequality is $1/2$.
In our setting for $l=4$ parties we can also compute the function $\text{NAND}_2(x)$ with the following linear map to generate the four inputs to the parties:
\begin{equation}
\left(\begin{array}{c} s_1\\s_2\\s_3\\s_4\end{array} \right) = \begin{pmatrix}
1 & 0 \\ 0 & 1 \\ 1 & 1 \\ 0 & 0  
\end{pmatrix}\left(\begin{array}{c} x_1\\x_2\end{array} \right)_{\oplus}.
\end{equation}
Now we can modify the set-up so that the third and fourth parties measure $Y$ for $s_{j}=0$, and $X$ for $s_{j}=1$. Equivalently, we could have applied a NOT to the values of $s_{j}$, as is described in the main paper. Thus the inequality in Eq. \ref{eq:beta4} can be rewritten as:
\begin{equation}
\frac{1}{4}\langle -XXYY-XYXY-YXXY+YYYY\rangle \leq 1
\end{equation}
The bound on the right-hand-side is attained with the GHZ state $(|0,0,0,0\rangle+|1,1,1,1\rangle)/\sqrt{2}$, as above; this bound is also the maximum allowed algebraically.
\subsection{Bounds on Three-Qubit Entanglement}
Here we derive bounds on the inequalities above when the four parties are limited to sharing tri-partite quantum resources. In this case three of the four parties can share a quantum state and the fourth party's outcome is local, i.e. determined by the measurement choice $s_{j}$ and shared randomness between the parties. Therefore, for each inequality, we can choose the party whose outcome $m_{j}$ will be a deterministic function $(-1)^{r_{j}s_{j}}$, where $r_{j}\in\{0,1\}$. Then we can optimise over all measurements and states for the other three parties with the same pre-processing as outlined above for each function.

To summarise, for all the functions above, and pre-processing outlined above, we can limit one of the four parties to having a deterministic outcome in the inequalities. In addition to this, we can adapt the numerical techniques in \cite{Werner2001,Zukowski2002}, to find the optimal violation for the remaining three parties. The optimal violation for the remaining three parties will be attained by a three-qubit GHZ state. This optimisation can be done for each choice of the party that does not share quantum resources with the other three.

For the function $f(x) = x_1 (x_2 \oplus x_3 \oplus 1) \oplus x_2 (x_3 \oplus 1) \oplus x_3$, the bound on $\beta$ for the inequality (\ref{eq:beta1}) for tripartite quantum resources is $\beta\leq 1/\sqrt{2}$. Notably this bound relates to the maximal quantum violation of the Svetlichny inequality \cite{collins2002}, as discussed in \cite{Hoban2011}. For the function $\text{OR}_3(x)$, the bound on $\beta$ for (\ref{eq:beta2}) for tripartite quantum resources is $\beta\leq 2/3$. For inequality  (\ref{eq:beta3}), the bound $\beta\leq0.70235$ holds. For the function $\text{NAND}_2(x)$, the function can be computed deterministically by three qubits, as shown in \cite{Anders2009}. 


\section{Further experimental details and data analysis}
 
For the experiment as illustrated in Fig.~3 of the main manuscript a 4\,W fs-pulsed Ti:sapphire laser at 780~nm is first frequency doubled and then directed onto a nonlinear $\beta$-barium-borate (BBO) crystal which produces spatially separated single photons by type-II spontaneous parametric down-conversion (SPDC). The two BBO crystals produce a polarization entangled state
\begin{equation}
\ket{\psi^{(\varphi)}} = \left(\ket{H,V}+e^{i\varphi}\ket{V,H}\right)/\sqrt{2}~.
\label{eq:BellPhaseState}
\end{equation}
\begin{figure}
	\centering
	\includegraphics[width=.94\textwidth]{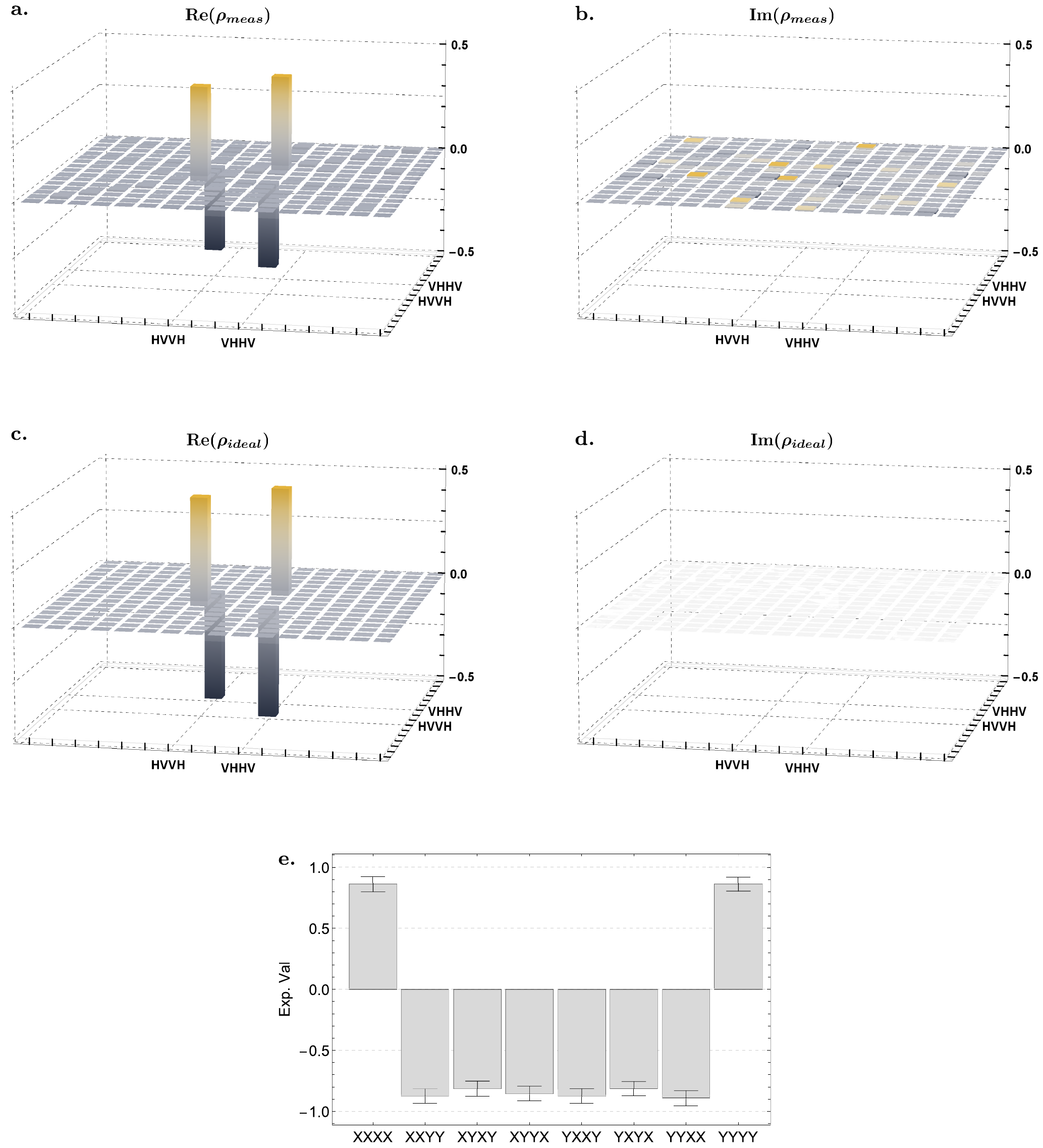}
	\caption{Panels \textbf{a-d.} show the measured real and imaginary part of the quantum state~$\rho_{exp}$ calculated by a maximum likelihood estimation from the measured 4-fold coincidence counts as well as their ideal cases. \textbf{e.} Calculated expected values obtained by the probability measurements of $4^2 = 16$ combinations to measure H or V polarized photons. The data was measured at 150\,mW pump power.}
	\label{fig:Tomos}
\end{figure}
After the entangled photon pairs are emitted in conic sections the photons in each spatial mode pass a combination of half-wave plate (HWP) 
and another BBO-crystal which compensate previously induced walk-off effects and allow to adjust the relative angle $\varphi$ in \cref{eq:BellPhaseState}. By fixing $\varphi = \pi$, a product of two Bell-states $\ket{\psi_{12}^{-}} \ket{\psi_{34}^{-}}$ is obtained, which upon superposition of modes 2 and 3 on a polarizing beam splitter (PBS) becomes
\begin{equation}
\text{PBS}_{23} \ket{\psi_{12}^{-}} \ket{\psi_{34}^{-}} = \frac{1}{\sqrt{2}}\left(\ket{\text{GHZ}'} + \ket{\chi} \right)~.
\end{equation}
The final state is thus a superposition of a four photon GHZ state $\ket{\text{GHZ}'} = \left(\ket{H,V,V,H} - \ket{V,H,H,V}\right)/\sqrt{2}$ and $\ket{\chi}= i\left(\ket{H,HV,0,H} + \ket{V,0,HV,V}\right)/\sqrt{2}$\,. Any 4-fold coincidence measured at the 8 avalanche photodiodes (APDs) is thus only obtained by the GHZ state.

To check the quality of the experimental state $\rho_{exp}$ we conducted a regular quantum state tomography by recording the $3^4$ combinations of the expected values $\text{Tr}(O_1 O_2 O_3 O_4 \rho_{exp})$, $O_i = \{X_i, Y_i, Z_i\}$ being the Pauli operators. To calculate a state fidelity we first get an estimation of the distribution~$\rho_{exp}$ from the measured data by maximizing a likelihood function and then compare to the optimal state~$F = \braket{\text{GHZ}'\vert \rho_{exp} \vert \text{GHZ}' }$. The data was obtained at 500\,mW pump power with a maximal 4-fold coincidence rate of 
ca.~1\,Hz in the eigenbasis of the projectors.
We obtained a value of~$F = 0.824 \pm 0.014$ with the the state tomography shown in~\cref{fig:Tomos}. The error has been estimated from a Monte Carlo simulation, i.e. random sampling from a Poisson distribution and iterating the fidelity calculation 100 times.
The expected values $E(x)$ that were recorded for the violation of the respective inequalities are exhibited in \cref{fig:Tomos}\,e.

\end{document}